# Interstellar Pickup Ion Observations to 38 AU

D.J. McComas[1,2], E.J. Zirnstein[1], M. Bzowski[3], H.A. Elliott[2], B. Randol[4], N.A. Schwadron[5], J.M. Sokół[3], J.R. Szalay[1], C. Olkin[6], J. Spencer[6], A. Stern[6], H. Weaver[7]

[1]Department of Astrophysical Sciences, Princeton University, Princeton, NJ 08544, USA
    (dmccomas@princeton.edu)
[2]Southwest Research Institute, San Antonio, TX 78228, USA
[3]Space Research Centre of the Polish Academy of Sciences, Warsaw, Poland
[4]Goddard Space Flight Center, Greenbelt, MD 20771, USA
[5]University of New Hampshire, Space Science Center, Durham, NH 03824, USA
[6]Southwest Research Institute, Boulder, CO 80302, USA
[7]Johns Hopkins University Applied Physics Laboratory, Laurel, MD 20723, USA

**ABSTRACT**
This study provides the first direct observations of interstellar $H^+$ and $He^+$ pickup ions in the solar wind from 22 AU to beyond 38 AU. We use the functional form of the Vasyliunas and Siscoe model, including pickup ion isotropization, convection, and adiabatic cooling, to fit to SWAP pickup ion observations. We are able to fit most observed distributions, although the fit parameters generally lie outside their physically expected ranges. None-the-less, this functional form allows fits that quantify the pickup $H^+$ density, temperature, and pressure over this range of distances. By ~20 AU, the pickup ions already provide the dominant internal pressure in the solar wind. We determine the radial trends of the pickup $H^+$ parameters, and extrapolate them to the termination shock at ~90 AU. There, the inferred ratio of the pickup $H^+$ to core solar wind density of ~0.14. We find that the pickup $H^+$ temperature and thermal pressure increase with distance over 22-38 AU, which indicates that there is likely additional heating of the pickup ions. This produces very large extrapolated ratios of pickup $H^+$ to solar wind temperature (~2400) and pressure (~350) at the termination shock. Similarly, the extrapolated ratio of the pickup ion pressure to the solar wind dynamic pressure at the termination shock is ~0.16. This value is over two and a half times larger than what is expected, based on a simple mass, momentum, and energy conservation model; if the ratio is anything close to this large, it has profound implications for moderating the termination shock and the overall outer heliospheric interaction. We also identify suprathermal tails in the $H^+$ spectra and several complex features in the $He^+$ pickup ion spectra, likely indicating variations in the pickup ion history and processing. Finally, we discover enhancements in both $H^+$ and $He^+$ populations just below their cutoff energies, which may be associated with enhanced local pickup. Altogether, this study serves to document the release and as the citable reference of these pickup ion data for broad community use and analysis.

*Key words:* Sun: heliosphere - (Sun) solar wind - ISM: general - (Galaxy:) local interstellar matter

## 1. INTRODUCTION

Neutral atoms in the very local interstellar medium (VLISM) move at a velocity ($\mathbf{v}_{LISM}$) with respect to the heliosphere of ~25.4 km s$^{-1}$ (e.g., McComas et al. 2015). When neutrals that enter the heliosphere are ionized, they immediately begin to respond to the motional electric field produced by the solar wind flow and its embedded interplanetary magnetic field (IMF). These newly created ions gyrate about the IMF and become incorporated into the solar wind as pickup ions. This process produces a velocity distribution quite distinct from the core solar wind, with pickup ions traveling at all speeds from nearly zero (their initial neutral flow speed) to twice the solar wind speed in the Sun's frame of reference, depending on the phase of their gyromotion. Pitch angle scattering quickly redistributes these ions onto a nearly-isotropic shell in velocity space, so they can be seen coming from all directions in *in situ* spacecraft measurements. Much more slowly they adiabatically cool, eventually filling in the shell with older pickup ions as new pickup ions are added to the outer shell of the distribution.

An early model of pickup ion velocity distributions was provided by Vasyliunas & Siscoe (V&S; 1976) and included ionization of the neutral interstellar gas, instantaneous scattering of the pickup ions into an isotropic distribution in the solar wind frame, convection, and adiabatic cooling. Early expectations for pickup ion distributions (Lee & Ip 1987) were that they would be roughly isotropic due to pitch angle scattering from both background turbulence and self-generated waves. The first *in situ* detection of interstellar pickup ions (Möbius et al. 1985) were based on observations from SULEICA (Supra-thermal Energy Ionic Charge Analyzer) on AMPTE (Active Magnetospheric Particle Tracer Explorer) and observations from the Solar Wind Ion Composition Spectrometer (SWICS) instrument on Ulysses (Gloeckler et al. 1992), which showed highly anisotropic distributions (Gloeckler et al. 1995). These observations indicated longer (~AU) scattering mean free paths, possibly owing to the strong two-dimensional component to turbulence in the solar wind (Matthaeus et al. 1990; Bieber et al. 1996) that is thought to be ineffective for pickup ion scattering (Bieber et al. 1994; Zank et al. 1998). Subsequent pickup ion models (e.g., Isenberg 1997; Schwadron 1998) have been developed to accommodate longer scattering mean free paths.

SWICS provided a broad range of pickup ion observations spanning the heliocentric distances sampled by the *Ulysses* mission from ~1.4-5.4 AU (see the review by Gloeckler & Geiss 1998 and references therein). These observations included pickup species of H$^+$, He$^+$, N$^+$, O$^+$, and Ne$^+$ (Geiss et al. 1994) as well as He$^{++}$ and $^3$He$^+$ (Gloeckler et al. 1997).

As the solar wind propagates out through the heliosphere, it incorporates increasingly more pickup ions (predominantly H$^+$ as this is the primary interstellar neutral species). As pickup ions continue to be incorporated into the solar wind, they progressively slow the wind and convert energy extracted from the solar wind bulk motion into pickup ion particle pressure. Effects of the pickup ion pressure have been studied by multiple authors (e.g., Fahr & Fichtner 1995, Lee 1998, Fahr & Scherer 2005). This pressure is the dominant internal pressure in the solar wind in the outer heliosphere and ultimately is expected to become a non-negligible fraction of the solar wind's overall dynamic pressure by the time the solar wind reaches the termination shock at ~100 AU. Thus, interstellar pickup ions incorporated into the solar wind play a critical role in the overall interaction of the heliosphere with the VLISM.

Because they extend to higher energies, these ions are more readily energized by a variety of acceleration processes and preferentially act as seed particles for energetic particle populations (e.g., Fisk & Lee, 1980, Schwadron et al. 1996, Chalov 2001, Giacalone et al. 2002, Fisk & Gloeckler, 2006, 2007, 2008; Chen et al. 2015). It is thought that pickup ions also significantly weaken the termination shock and receive most of the energy available at the shock instead of the core solar wind (Richardson et al. 2008). Beyond the termination shock, they provide a significant fraction of the pressure inside the heliopause in the inner heliosheath, as shown through Energetic Neutral Atom (ENA) observations (e.g. Livadiotis et al. 2013) from the Interstellar Boundary Explorer (IBEX) mission (McComas et al. 2009a, 2009b).

Prior to the Solar Wind Around Pluto (SWAP) instrument (McComas et al. 2008) measurements on the *New Horizons* spacecraft, there was very little direct information about pickup ions beyond Jupiter's orbit (*Ulysses'* perihelion). McComas et al. (2004) used *Cassini* spacecraft data from 6.4-8.2 AU in the downwind direction from the inflowing interstellar neutrals to 1) show enhancements in $He^+$ from gravitational focusing by the Sun (Thomas 1978; Weller & Meier 1981; Möbius et al. 1985) and 2) make the first in situ observations of depletion of $H^+$ pickup ions in a downwind "interstellar hydrogen shadow." Indirect observations from Pioneer 10 at ~8.3 AU (Intriligator et al. 1996) found Doppler-shifted ion cyclotron waves possibly generated by $H^+$ pickup ions and Mihalov & Gazis (1998) claimed "possible signatures of interstellar pickup hydrogen" to 16 AU in Pioneer 10 and 11 data.

The SWAP instrument was designed to measure the solar wind and pickup ions out at ~33 AU, as the *New Horizons* spacecraft repeatedly rotated to point its various cameras during the flyby of Pluto and Charon (Stern et al. 2015). Because of the great distance and large range of viewing directions, SWAP was designed to have an extremely high sensitivity and a very large field-of-view (FOV). These attributes allowed SWAP to make fundamental measurements of the jovian magnetosphere and magnetotail (McComas et al. 2007, 2017; Ebert et al. 2010; Nicolaou et al. 2014, 2015a, 2015b) and Pluto's interaction with the solar wind (McComas et al. 2016; Bagenal et al. 2016; Zirnstein et al. 2016). See McComas et al. (2008) for details of the SWAP instrument.

SWAP is ideally suited to make high quality observations of interstellar pickup $H^+$, and has extended these detailed pickup ion observations beyond 8.2 AU for the first time (McComas et al. 2010). Internally generated backgrounds inside the sensor were characterized by Randol et al. (2010) based on computer simulations of particle trajectories and scattering inside instrument, which provided even higher quality pickup ion observations from SWAP. In two subsequent studies, Randol et al. (2012, 2013) extended the sparse SWAP pickup ion observations from 11 AU out to ~17 AU and ~22 AU, respectively. These authors compared the pickup ion spectra to the classic V&S model and showed reasonable agreement when using ionization rates consistent with independently derived averages and an increase over time with increasing solar activity.

New Horizons is headed only ~30° from the interstellar inflow direction. This makes SWAP observation of pickup ions much easier to understand than if the spacecraft were headed at a large angle from this direction. This is because the interstellar neutrals that produce the pickup ions observed by SWAP have traveled nearly radially inward from the upstream interstellar medium as shown in Figure 1.

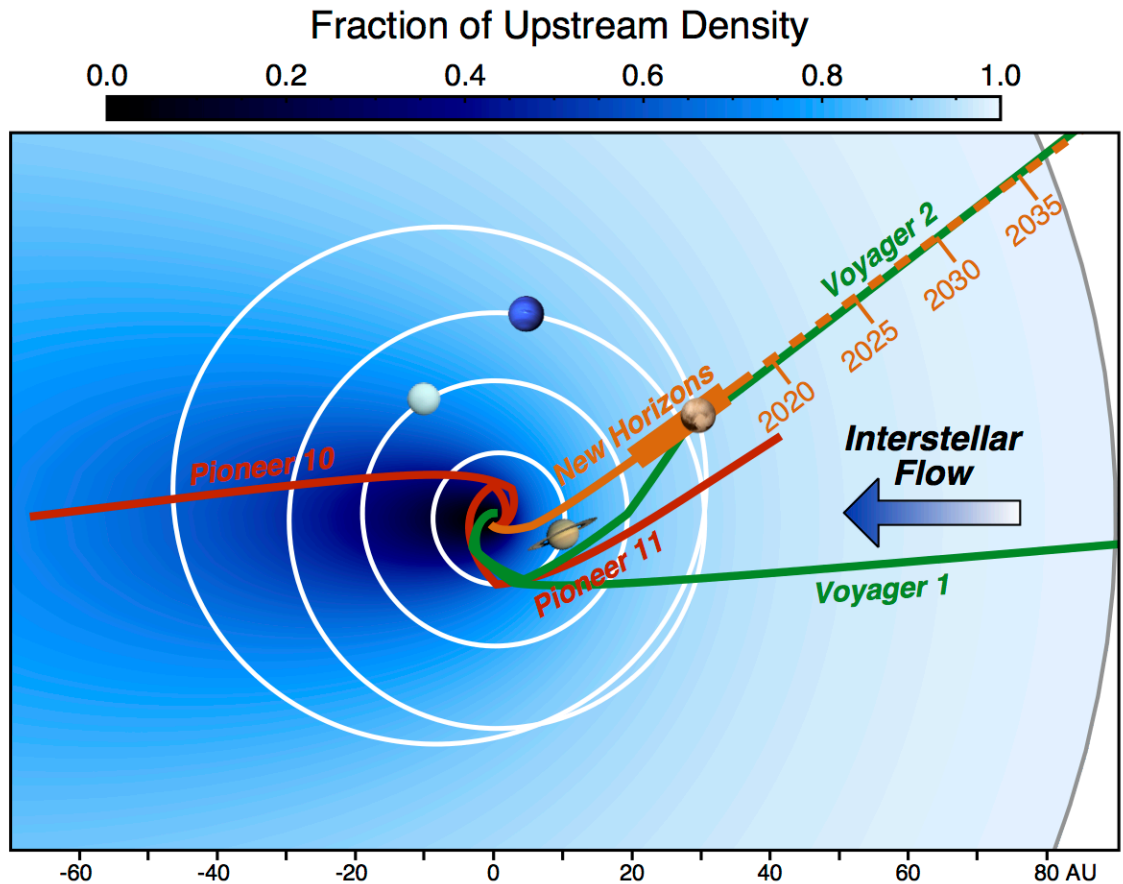

*Figure 1. Schematic diagram of the New Horizons trajectory (orange) compared to the Voyager (green) and Pioneer (red) trajectories, all projected into the ecliptic plane (Voyager 1 and 2 are ~30° above and below this plane, respectively). The colored background indicates the fraction of the upstream interstellar H that survives at various locations from the "hot model" (Thomas 1978; Wu & Judge 1979) as was done to show*

*the interstellar hydrogen shadow (dark region downwind of the Sun) by McComas et al. (2004).*

For earlier SWAP studies (McComas et al. 2010; Randol et al. 2012, 2013), the instrument was turned on only for brief intervals and no detailed statistical analysis was possible. However, since 2012, when *New Horizons* was at ~22 AU, SWAP has been left on the majority of the time, even during extended spacecraft "hibernation" intervals. This has allowed us to produce a much more nearly continuous data set of outer heliospheric solar wind observations. Elliott et al. (2016) developed the techniques and published core solar wind observations from SWAP out to ~33 AU.

In this study, we examine the extensive set of SWAP pickup ion observations from ~22 AU out to beyond 38 AU for the first time. We also provide the citable reference and the detailed documentation for the release of these data to the community so that they can be used by other interested researchers. Importantly, these data cover the range of heliocentric distances where the pickup ion pressure is expected to become the dominant internal pressure in the solar wind, exceeding the thermal pressure of the original solar wind particles and magnetic pressure of the IMF. These observations from SWAP enable key new insights into the critical role of interstellar pickup ions in the outer heliosphere and its interaction with the VLISM.

## 2. INTERSTELLAR PICKUP ION SPECTRA

The SWAP instrument on New Horizons utilizes a top-hat electrostatic analyzer (ESA), with a large field-of-view (FOV; 276° x 10°). SWAP detects ions in 64 logarithmically-spaced energy per charge (E/q) bins covering 0.023-7.87 keV/q, with an energy resolution $\Delta E/E$ of 8.5% FWFM (McComas et al. 2008). SWAP steps through the range of E/q bins in 64 s, first a "coarse" sweep over the whole energy range in the first 32 s, then a "fine" sweep over a narrower energy range at higher resolution, centered on the peak of the solar wind distribution. Each E/q step has a sampling time of 0.39 s.

Ions that are accepted through the ESA over a specific range of E/q are then electrostatically accelerated into an ultra-thin carbon foil (McComas et al. 2004), where electrons are ejected. Primary ions and electrons from the exit surface of the foil are detected as signals by the primary channel electron multipliers (PCEM) and electrons from the entrance (front) surface of the foil are steered into and detected in a secondary CEM (SCEM). A coincidence (COIN) count is generated if both PCEM and SCEM detections are made within a 100 ns window. COIN measurements provide excellent signal-to-noise measurements of the solar wind (SW) energy spectra (McComas et al. 2008), solar wind bulk parameters (Elliott et al. 2016), and interstellar pickup ions (McComas et al. 2010; Randol & McComas, 2010; Randol et al. 2012, 2013).

Figure 2 shows an example of SWAP observations at ~25.7 AU. The coincidence data in this spectrum were accumulated over a 24 hr period. The solar wind protons peak at ~600 eV/q in this example, while the $He^{++}$ peaks at twice the E/q (~1200 eV/q). The other counts at lower and higher energies are produced by a combination of noise/background, and interstellar $H^+$ pickup ions up to a cutoff at four times the solar wind energy. Above the $H^+$ cutoff, the interstellar $He^+$ pickup ions take over and generally produce the relatively flat ledge seen out to the highest E/q bins measured. This ledge was not identified as being $He^+$ pickup ions in previous SWAP studies (McComas et al. 2010; Randol et al. 2012, 2013), so the current study is the first to make this identification. Because the maximum E/q measured is 7.8 keV/q, SWAP does not observe the $He^+$ cutoff at four times the energy of the $H^+$ cutoff (16 times the solar wind E/q), although, when the solar wind is very slow, the observations come close to this value.

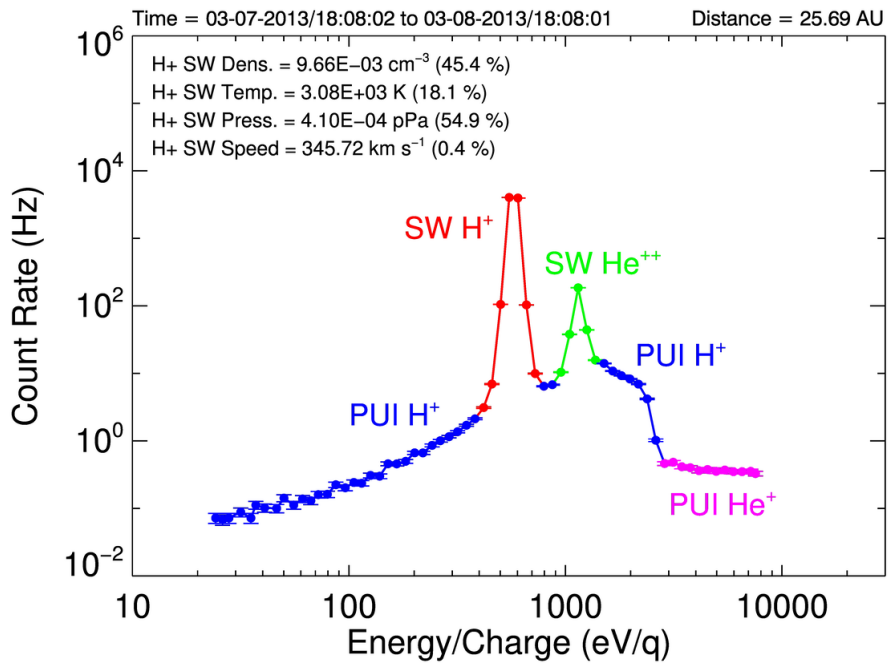

*Figure 2. Example of SWAP observations taken at ~25.7 AU, colored-coded and annotated with the primary source of the counts: solar wind (SW) or interstellar pickup ion (PUI). One-day averages collected on the spacecraft provide highly statistically significant measurements (error bars are too small to be seen except at the lowest E/q). The average of the hourly solar wind bulk parameters derived over this day (Elliott et al. 2016) are shown in the upper left of the figure. Percentages in brackets on the solar wind parameters indicate the normalized root mean square (RMS) variations of the hourly samples.*

As described above, the V&S model includes ionization of the neutral interstellar gas, instantaneous scattering of the pickup ions into an isotropic shell distribution in the solar wind frame, convection, and adiabatic cooling. For completeness, here we derive the pickup ion distribution, starting with the transport equation

$$u_p \frac{df}{dr} - \frac{2u_p}{3r} w \frac{df}{dw} = N\beta(r) exp\left(-\frac{\lambda}{r}\frac{\theta}{sin\theta}\right)\frac{\delta(1-w)}{4\pi u_p^3}, \quad (1)$$

which includes the convection of pickup ions with the solar wind (first term), adiabatic cooling due to the radial expansion of the solar wind (second term), and the pickup ion source term, which accounts for the creation of pickup ions and assumes gravity and radiation pressure are balanced (right side). Equation (1) includes the radial and angular dependence of the interstellar neutral atom (H or He) density, given by the exponential term. For simplicity, here we use a form for the neutral density that assumes a neutral temperature at infinity of 0 (i.e., the cold model), and that the effects of gravity and radiation pressure are balanced (i.e., $\mu = 1$). Solving Equation (1) gives the pickup ion distribution as a function of distance, $r$, from the Sun

$$f(r,w) = \frac{3N\beta_E r_E^2 \Theta(1-w)}{8\pi u_p \tilde{u}^3 r w^{\frac{3}{2}}} exp\left(-\frac{\lambda}{rw^{\frac{3}{2}}}\frac{\theta}{sin\theta}\right), \quad (2)$$

where $N$ is the interstellar neutral density far from the Sun, $\beta$ is the local ionization rate (such that $\beta(r) = \beta_E (r_E/r)^2$), $u_p$ is the bulk solar wind flow speed, $\tilde{u} = u_p + v_{LISM} \cos(\theta)$, $w$ is the ratio of the pickup ion particle speed ($v$) to the injection speed such that $w = v/\tilde{u}$, $v_{LISM}$ is the interstellar flow speed, $\lambda$ is the interstellar neutral ionization cavity length, and $\theta$ is the angle between the radial vector to SWAP's position and the VLISM inflow direction (ecliptic longitude 255.7°, latitude 5.1°; McComas et al. 2015). We note that in the derivation of Equation (2), we introduce the small, but finite, relative velocity between the radially propagating solar wind and the inflowing LISM into the PUI injection/cutoff speed. This is achieved by replacing $u_p$ on the right-hand side of Equation (1) with $\tilde{u}$, and setting $w = v/\tilde{u}$. The pickup ion distribution has a maximum speed of $v = \tilde{u} \sim u_p$, or $w = 1$, in the plasma frame. Thus, Equation (2) includes a Heaviside step function $\Theta(1-w)$.

The model assumes the source ($\beta$) and loss (related to $\lambda$ by $\beta_L = \lambda v_{LISM}/r_E^2$) ionization rates are not instantaneously balanced. Moreover, a different value for $\mu$ could alter the values of both $\beta$ and $\lambda$ by altering the density inside and around the ionization cavity, which maps to velocity space distribution functions even at 20-40 AU (see discussions by Mobius et al. 1988; Ruciñski & Bzowski 1995, 1996; Ruciñski et al. 2003).

To model SWAP observations, we follow Randol et al. (2013), and the pickup ion distributions are converted to count rates

$$C(v) = \frac{gv^4}{\Delta\phi} \int_{-\Delta\phi/2}^{\Delta\phi/2} f(v,\phi)d\phi \quad (3)$$

where $\varphi$ is the azimuthal angle of SWAP's FOV ($\Delta\varphi = 276°$), $f$ is the pickup ion distribution from Equation (2), and $g$ is SWAP's energy-dependent geometric factor. The geometric factor is 2.1 x $10^{-3}$ cm$^2$ sr eV/eV at 1 keV/q, and varies with energy (especially at lower energies) due to post-acceleration of ions exiting the ESA (McComas et al. 2008; Randol et al. 2012).

First turning to interstellar He pickup ions, we apply Equation (2) to interstellar $He^+$ pickup ions in the solar wind, taking the interstellar neutral He density far from the Sun, $N_{He} \sim 0.015$ cm$^{-3}$ (Witte 2004, Gloeckler et al. 2004). For the He ionization cavity length, $\lambda_{He}$, we use a fixed value for all spectra of 0.5 AU from. The ionization rates of neutral He determines the production of $He^+$ pickup ions in the solar wind. At one AU, photoionization is the dominant ionization process, being an order of magnitude larger than electron-impact ionization and two orders greater than charge-exchange (e.g., Bzowski et al. 2013). At large distances from the Sun, the rate of photoionization drops as $r^{-2}$, but electron-impact ionization drops even more quickly, so photoionization is the only significant ionization source for the outer heliospheric $He^+$ pickup ions examined here; thus, we only include photoionization of He. Finally, this ionization rate evolves over the 11-year solar cycle. Here we spline interpolate daily values from the Carrington rotation-averaged He photoionization rate (Figure 3), derived from measured solar spectra and solar EUV radiation proxies (Bzowski et al. 2013; Sokół & Bzowski 2014).

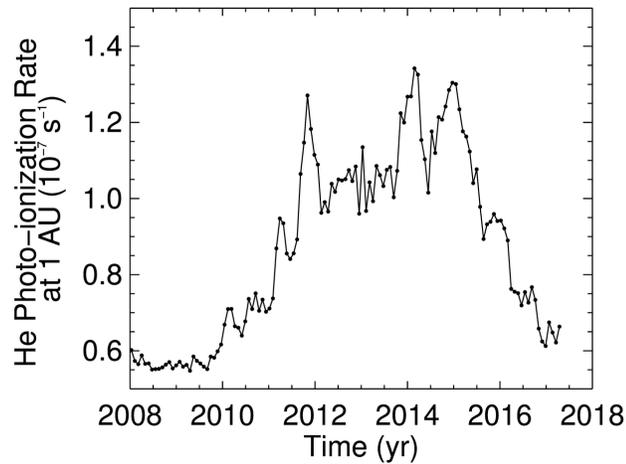

*Figure 3. Carrington rotation averaged He photoionization rate at 1 AU.*

This same approach of applying the V&S model is also used for the SWAP observations of interstellar $H^+$ pickup ions in the solar wind (Randol et al. 2012, 2013). However, instead of using external values for the ionization rate ($\beta$) and cavity length ($\lambda$), here we allow $\lambda$ and $\beta$ to be free fitting parameters. In Figure 4, we show the results of calculating the $He^+$ pickup ion distribution (orange) as described above, and fitting Equations (2) and (3) to the $H^+$ pickup ion observations (blue). After considerable experimentation with fits to various portions of the $H^+$ distribution, we settled on fitting simultaneously over two portions of this distribution (including their individual uncertainties): 1) energy steps less than one half of the E/q of the highest data point in the solar wind peak and 2) the three E/q steps just below the pickup ion cutoff. The cutoff is calculated using both the solar wind speed and much smaller VLISM flow speed, so it is at about four times the bulk solar wind peak energy (note: we also remove from the fit the highest of these three E/q steps if within 2% of the model cut-off because this extremely sharp feature overly biases the fit for points that are too close to it).

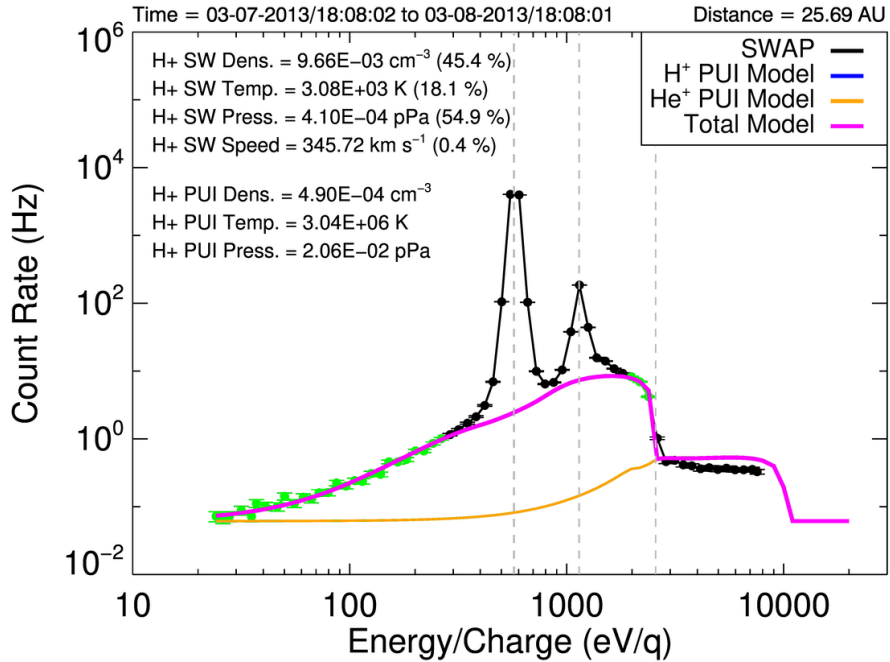

*Figure 4. Model results compared to the SWAP pickup ion observations from Figure 2. Vertical dashed lines indicating the peak locations of the solar wind (i.e., solar wind bulk flow speed in SWAP's frame), $He^{++}$, and the energy cutoff of $H^+$ pickup ions at twice the solar wind speed (four times the energy). The fit for $H^+$ included points (green) that cover the low energy range and just below the pickup ion cutoff. The model curve for $He^+$ (orange) is far simpler, and depends only on the variation in the He photoionization rate, so significant differences from the data are not unexpected.*

The limit at half of the solar wind E/q is driven by backgrounds from internal scattering of ions from the intense solar wind beam. Scattering off the inside surface of the outer electrostatic analyzer plate in the SWAP instrument can contribute significantly at energies above this value (Randol et al. 2010). We further limit the fitting to the 2-3 points just below the cutoff energy to avoid "contamination" from: 1) the alpha particle distribution at times when the solar wind is very hot and 2) solar wind heavy ions, such as $O^{+6}$ at 2.7 times the solar wind E/q, which can be enhanced, often associated with CMEs. Finally, the fitting is an iterative chi-squared minimization process to ensure that it finds the best overall fit globally, we start the process with a variety of initial parameters spanning the range of likely physical values and chose the solution with the lowest overall chi-squared difference from the observations.

For the spectrum shown in Figure 4, the measured solar wind speed is 346 km s$^{-1}$ and the externally provided, one AU scaled He ionization rate at this time, $\beta_{He} \times (r/r_E)^2$ is $1.04 \times 10^{-7}$ s$^{-1}$. We assume a nominal interstellar neutral H density ($N_H$) of 0.1 cm$^{-3}$ (Bzowski et al. 2009 find 0.09+/-0.022). Iteratively running the model for H returns a one AU scaled H ionization rate $\beta_{HF} \times (r/r_E)^2$ of $5.63 \times 10^{-7}$ s$^{-1}$ and a H ionization cavity length $\lambda_{HF}$ of 6.24 AU. We use the subscript "HF" here to note that these are fit parameter results derived for the specific H pickup ion distribution and are not taken from separate data or other considerations as model inputs. Finally, the fitting returns an energy-independent background count rate, which is largely believed to originate from galactic cosmic rays.

For this spectrum, the background count rate is ~0.0612 Hz. This energy-dependent background count rate is the reason for the flat level count rate at highest and lowest E/q.

The magenta curve Figure 4 indicates the values from the overall fitting at each of the E/q steps used, and shows excellent consistency with the measured spectrum. The overall reduced chi-squared for the H portions of the fit is 4.31; this difference is largely driven by the points just below the cutoff, where the values are largest and thus small relative differences can still be large. The reduced chi-squared for the He ledge, which is modeled but not fitted (see above), is calculated using all E/q steps above 1.2 time the $H^+$ cutoff energy, and is 34.01 for this example. Because the He model uses a fixed $\lambda_{He}$ of 0.5 AU and interpolates the Carrington rotation UV observations to fix each day's $\beta_{He}$, the differences can be significant.

Figure 5 provides three samples of the solar wind and pickup ion distributions from 2/21 (black), 3/1 (red), and 3/6 (blue) 2013, when the spacecraft was at approximately 25.6 AU. As shown in Figure 6, the spectrum starting on 3/1 is from a slightly faster and significantly denser (~80%) and hotter (~60%) region, which likely represents solar wind that was in a large compression region or interplanetary shock closer in to the Sun. This spectrum also indicates a significant enhancement in more recently picked up ions (seen in the red curve at energies above the alpha particles) to $8.2 \times 10^{-4}$ $cm^{-3}$ (from $4.8 \times 10^{-4}$ $cm^{-3}$ on 2/21). Six days later, on 3/6/2013, this enhancement at higher energies had been reduced and was similar to the 2/21/2013 shape, but the "older" portion of the pickup ion distribution just below the solar wind peak was even more enhanced than in either of the prior times. This sequence of observations is consistent with both the enhanced pickup of ions by denser solar wind and the processing of pickup ions to lower energies as they cool over time.

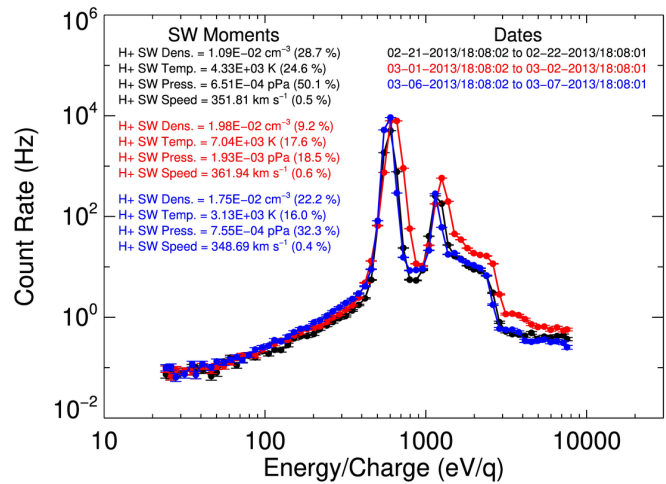

**Figure 5.** *SWAP E/q spectra for three days in 2013: 2/21 (black), 3/1 (red), and 3/6 (blue).*

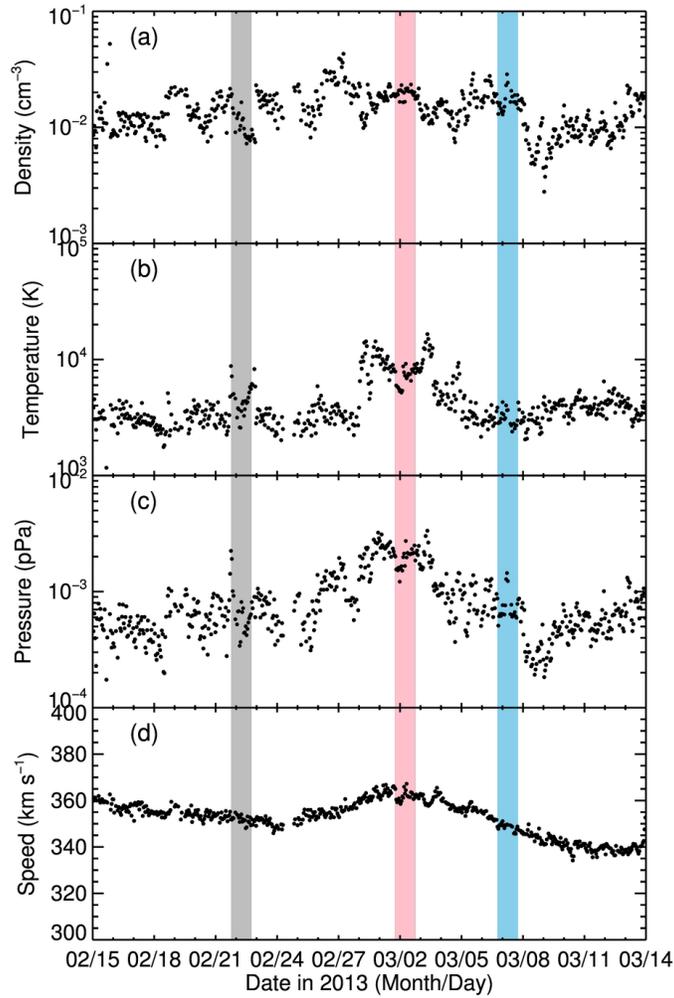

*Figure 6. Solar wind parameters taken by SWAP for the interval surrounding the spectra in Figure 5. Colored regions indicate the intervals for these spectra on 2/21 (grey), 3/1 (red), and 3/6 (blue).*

## 3. RADIAL TRENDS IN PICKUP HYDROGEN PARAMETERS

Over the six years interval where New Horizons transited from ~22 out to ~38 AU, we analyzed 1156 roughly one-day spectra. Each of the spectra was fit through the forward modeling and chi-squared minimization process described in Section 2, which included varying initial parameters to search a broad parameter space and ensure that the fitting process found the best fit globally. Then, we removed spectra where solar wind speed varied over the day by >1% (~13% of the samples). As a final quality control, each of the remaining spectra were visually compared to their V&S fit curve and spectra where fitting somehow failed (~2%) were manually removed. A few of these instances were inexplicable, while others indicated physical processes where the pickup ion spectra were simply unlike the V&S model shape; these and other interesting spectra are discussed in Section 5. The quantitative results for the remaining ~85% of the full collection of spectra are examined here.

Figure 7 shows plots of the solar wind and pickup $H^+$ parameters (density, temperature, and pressure) vs heliocentric distance. The solar wind values are calculated from the moments provided by Elliott et al. (2016) out to 33 AU and were extended using the same technique thereafter. The pickup ion parameters come from the fitting procedure developed in this study. A power law function was fit for each parameter and population, and the resulting fits are shown in Figure 7 as orange and green curves, for the solar wind and PUIs, respectively. We also binned the data into sidereal rotation periods (only plotted if there are at least 10 samples in that period) and provide the mean and +/- one sigma ranges (black/red points). Finally, as a cross-check, these binned data were also fit with a power law function and for the pickup ion $H^+$ pressure we calculated a radial γ of 0.22+/-0.26, consistent with the value of 0.10 from fitting the individual data points.

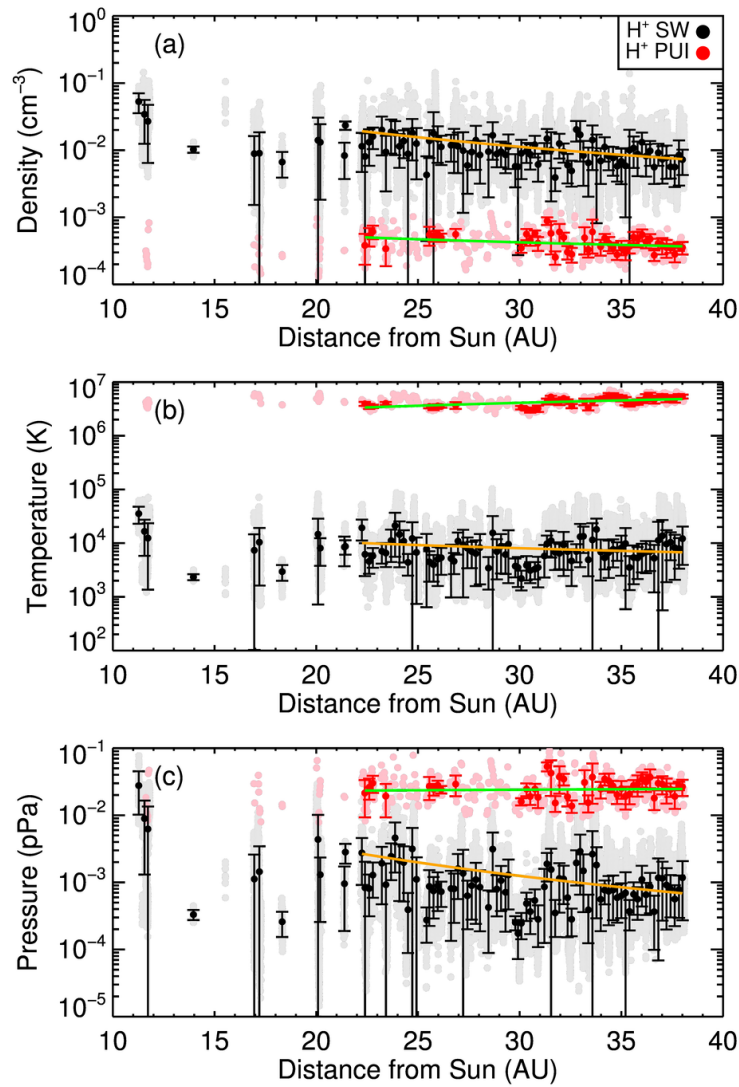

*Figure 7.* Densities, temperatures, and internal particle pressures (nkT) for solar wind ions (black) and $H^+$ pickup ions (red). Power law fits to all data (grey and pink points - not time-averaged data) show the radial variation in these parameters for the solar wind ions

*(orange) and pickup ions (green). Data time-averaged over sidereal periods, are over-plotted in red and black. Bars indicate +/- one sigma variations in each sidereal period (only plotted if there are at least 10 samples in that period). Note that we only fit to data beyond 22 AU where data coverage is large enough that the values are less biased by which portions of the solar wind happen to be sampled.*

We use power law fits to all non-culled data points before averaging and extrapolate each of these functions to the termination shock, assumed to be at ~90 AU near the upwind direction, as Voyager 1 crossed the termination shock at 94 AU (Stone et al. 2005) and Voyager 2 at 84 AU (Stone et al. 2008). Table 1 provides values from the fits in the middle of the range of the SWAP observations (30 AU), the radial dependences, and extrapolated values out at 90 AU. Ratios of the various values are indicated along the right-hand column and bottom two rows.

*Table 1. Solar wind (SW) and pickup ion (PUI) $H^+$ values from the fits at 30 AU and extrapolated to 90 AU along with the radial dependences ($r^\gamma$) and various ratios of these parameters.*

|  | Density (cm$^{-3}$) |  | Temp (K) |  | nkT Press (pPa) |  | Dyn Press (pPa)* | PUI Press / SW Dyn Press |
|---|---|---|---|---|---|---|---|---|
|  | SW | PU H+ | SW | PU H+ | SW | PU H+ | SW |  |
| 30 AU | 1.1x10$^{-2}$ | 4.2 x10$^{-4}$ | 8.0x10$^3$ | 4.1x10$^6$ | 1.2x10$^{-3}$ | 2.4x10$^{-2}$ | 1.4 | 0.017 |
| 90 AU | 1.6x10$^{-3}$ | 2.2 x10$^{-4}$ | 3.6x10$^3$ | 8.7x10$^6$ | 7.7x10$^{-5}$ | 2.7x10$^{-2}$ | 0.17 | 0.16 |
| Radial dep. ($\gamma$) | -1.8 | -0.58 | -0.74 | 0.68 | -2.5 | 0.10 |  |  |
| PUI/SW (30 AU) | 0.038 |  | 510 |  | 19 |  |  |  |
| PUI/SW (90 AU) | 0.14 |  | 2400 |  | 350 |  |  |  |

*The solar wind speed at 30 AU is on average ~390 km s$^{-1}$ (Elliott et al. 2016). By 90 AU, the solar wind speed decreases ~10-15% due to mass-loading of pickup ions; here we use 356 km s$^{-1}$ for the speed at ~90 AU (this speed is taken from a simple energy conservation model that includes pickup ion mass loading, described in the text). The dynamic pressure is calculated as $1/2 m_p n_p u_p^2$. The power law fits are of the form $Ar^\gamma$. We use the fixed values for the amplitudes ($A$) from Elliott et al. (2016) for the solar wind data, and allow $A$ to be a free fitting parameter for the pickup ion fits.

While there have been a variety of theoretical considerations of the properties of interstellar pickup ions in the heliosphere and out at the termination shock, Table 1 provides the first direct observational values beyond 22 AU. While the pickup density is still small compared to the solar wind density at 30 AU (~0.04), the much larger temperature (~500 times higher) means that the pickup particle pressure is already ~20 times that of the core solar wind by this distance. Though New Horizons does not carry a magnetometer, the magnetic pressure measured by Voyager at ~30 AU is also several times smaller (see Bagenal et al. 2015 for a summary of pressures at 33 AU), which means that the pickup ions are the dominant solar wind internal pressure (as opposed to dynamic pressure) by this distance. Figure 8 compares the radial trends for the various pressures observed by SWAP and compares them to other values from the literature. Because the pickup ion data are unique (blue line), they provide the first observational information about this key pressure at large (>20 AU) heliocentric distances, and its radial trend.

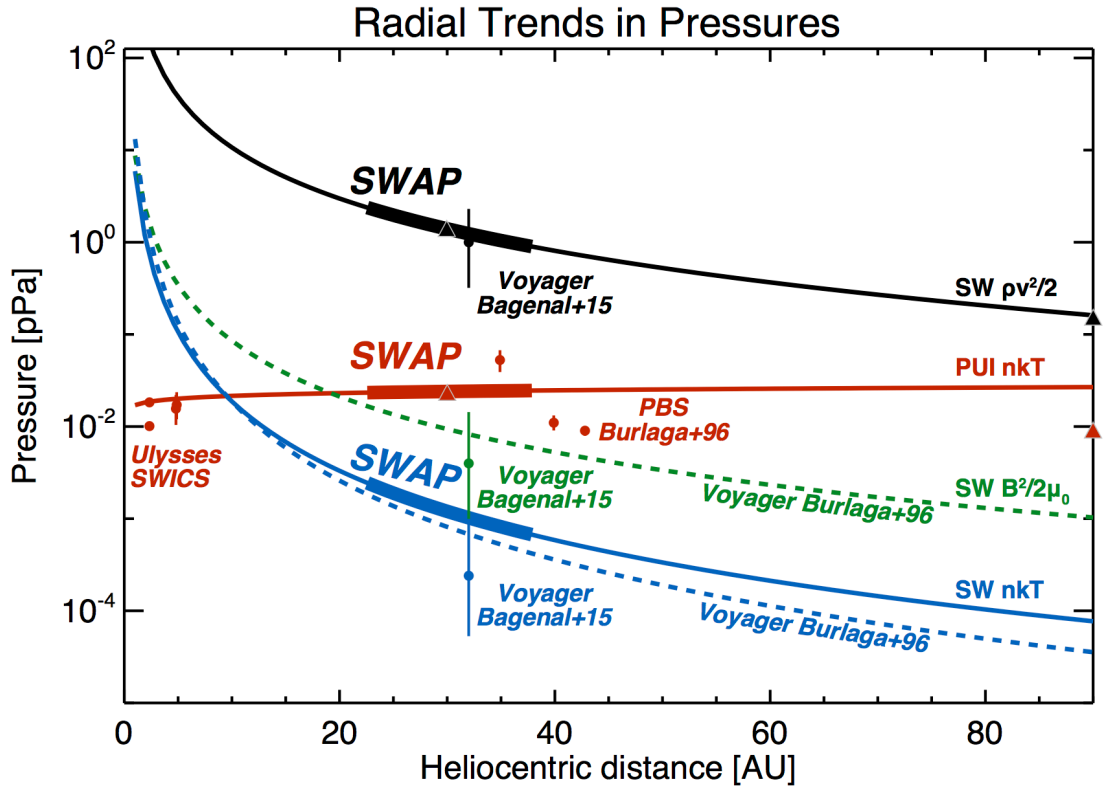

*Figure 8. SWAP observations (thick lines) from 22-38 AU and extrapolated radial trends (thin lines) of the solar wind (SW) dynamic pressure (black) and SW thermal pressure (blue) as well as the pickup $H^+$ (pickup ion) pressure (red). Magnetic pressure (green dashed) and another measure of solar wind particle pressure (blue dashed) are taken from Voyager (Burlaga et al. 1996) and averaged values and ranges at 33 AU come from Bagenal et al. (2015). Additional measurements of the pickup ion pressure from the Ulysses SWICS instrument out to 5 AU and inferred values from pressure balance structures (PBSs, Burlaga et al. 1996) are plotted in red. Finally, values from a simple conservation model (see text) are indicated with triangles in matching colors at 30 AU and the termination shock at 90 AU.*

The radial trends indicate that while the solar wind density is falling off roughly as $r^{-2}$, the pickup ion density is decreasing far less rapidly ($\sim r^{-0.6}$). This is because even though the volume of a parcel of plasma is expanding as $r^2$, the continued addition of pickup ions competes strongly with this reduction. Note that this radial dependence is different than the $r^{-1}$ dependence found from a first-order approximation to models of pickup ion mass-loading (e.g., Lee et al. 2009). Moreover, while the solar wind temperature and pressure are decreasing with distance (~$r^{-0.7}$ and ~$r^{-2.5}$, respectively), the pickup ion temperature and pressure are increasing (~$r^{+0.7}$ and ~$r^{+0.1}$, respectively). Further, we note that by including errors associated with the variation during sidereal rotations, we find that a pressure exponent of $r^{+0.22+/-0.26}$.

At the termination shock, the extrapolations indicate that pickup H+ should have a density ~0.14 and a particle pressure ~350 times that of the core solar wind. Further, because the pickup ion internal pressure is increasing over the range of observations from 22-38 AU, its extrapolated value is very large by the termination shock and is even a significant fraction of the solar wind's overall dynamic pressure (~0.16). Such a large fraction would moderate and have profound implications for the termination shock as indicated by the Voyager 2 crossing observations (Richardson et al. 2008).

We carried out this extrapolation two ways: 1) we fit the pickup $H^+$ density and temperature and the solar wind density and extrapolated them independently to the TS, and 2) we fit the pickup $H^+$ thermal pressure and solar wind density and extrapolated both to the TS. In both cases, we used the Elliott et al. (2016) value for the solar wind speed of ~390 km s$^{-1}$ at 30 AU and then assumed 356 km s$^{-1}$ at 90 AU, consistent with slowing of ~10-15% from mass loading. While the average of the product is not the product of the averages in general, here both approaches returned a pickup $H^+$ particle pressure to solar wind dynamic pressure ratio of 0.16. In addition, we checked to make sure that omitting the sparse SWAP pickup ion data from inside 22 AU didn't somehow bias the results, so we carried out the same two extrapolations, but including these data too. Both of those calculations returned an even larger ratio of pickup H particle to solar wind dynamic pressure ratio of 0.19. We prefer to use the 0.16 value instead of this larger one because the infrequent sampling of the solar wind in the earlier data could unfairly bias the results, whereas data from full (or large fractions) of solar rotations should sample all solar wind types.

As a cross check on the various extrapolated results, we modeled the internal pressure, mass flux, and bulk speed of the solar wind, including the creation of pickup protons, loss of solar wind protons through charge-exchange, and photoionization of interstellar neutral H using the methodology detailed in Schwadron et al. (2011). In this case, a steady radially expanding solar wind is mass-loaded due to the addition of pickup protons. The model has relatively few input parameters that strongly influence solutions over 20-90 AU: the mass flux and speed of solar wind on the inner boundary (taken to be 1 AU), the density of interstellar neutral H beyond the termination shock and the photoionization rate. Taking an inner boundary speed of 410 km/s, a mass flux of $3.9 \times 10^8$ cm$^{-2}$ s$^{-1}$ at 1 AU, a photoionization rate at 1 AU of $1 \times 10^{-7}$ s$^{-1}$ and an interstellar H density at the termination shock of 0.075 cm$^{-3}$, gives at 30 AU a pickup $H^+$ internal pressure of 0.024 pPa and density of $5.3 \times 10^{-4}$ cm$^{-3}$, and a solar wind bulk speed of 395 km/s and density of 0.011 cm$^{-3}$. These values are reasonably consistent with observations in Table 1 (see triangles in Figure 8). For example, the modeled ratio of the pickup ion density to solar wind density is 0.048, which is similar to the observed density ratio of 0.038 and modeled ratio of the internal plasma pressure to the ram pressure of 0.017 is the same as the observed pressure ratio.

Model results indicate that both the pickup ion density and pressure drop with distance, but do so more gradually than the solar wind density and pressure. By 90 AU, the modeled pickup H$^+$ internal pressure drops to 0.0095 pPa, solar wind ram pressure drops to 0.145 pPa, pickup ion density drops to 2.5x10$^{-4}$ cm$^{-3}$, and solar wind density drops to 1.4x10$^{-4}$ cm$^{-3}$. The speed of the solar wind also continues to drop as the plasma is mass loaded falling to 352 km s$^{-1}$ by 90 AU, constituting a 10% decrease in speed. Note that by 90 AU the modeled ratio of pickup ion density to the solar wind density is 0.18, similar to the extrapolated density ratio of 0.14.

The one significant difference between the extrapolated and model results at 90 AU is the ratio of the pickup ion internal pressure to solar wind ram pressure. The model returns a ratio of 0.65 at this distance, which is less than one third of the extrapolated pressure ratio of 0.16. This begs the question as to why the modeled pickup ion pressure, which assumes a steady radial expansion of the solar wind and conserves energy, mass and momentum, continues to fall-off with distance, whereas the extrapolated pickup ion pressure shows a small increase with distance.

One possibility for the observed increase in pickup ion pressure over 22-38 AU, and thus larger extrapolated value at 90 AU, is that some form of heating may be acting on the pickup ion population. For example, New Horizons is clearly measuring plasma in co-rotating interaction regions (CIRs), some of which may have become merged interaction regions (MIRs) by these distances; such plasma interaction and merging inherently involves some level of compression and heating. Further, it is likely that such compressions continue to act on the plasma, thereby increasing the internal pressure out to the termination shock.

Figure 9 shows the relation between the measured pickup ion density and local solar wind density. The general correlation of pickup ion to solar wind density, as reported before (Moebius et al. 2010; Randol & McComas, 2010; Randol et al. 2013), is not surprising both because 1) when a parcel of plasma compresses, the volume decreases for both the core solar wind and pickup ions, and 2) regions of higher solar wind density produce higher rates of charge exchange with the incoming interstellar neutrals and thus generates more pickup ions. Schwadron et al. (1999) showed enhancements of interstellar pickup H$^+$ and He$^+$ in high-latitude compressional regions in the solar wind using Ulysses observations. They explained these enhancements in terms of an ion transport model with long scattering mean free paths. Finally, because large-scale, higher-density, compression regions in the solar wind persist over many AU (even tens of AU) as they propagate outward, the effect of enhanced charge exchange is accumulated as such parcels of plasma move outward.

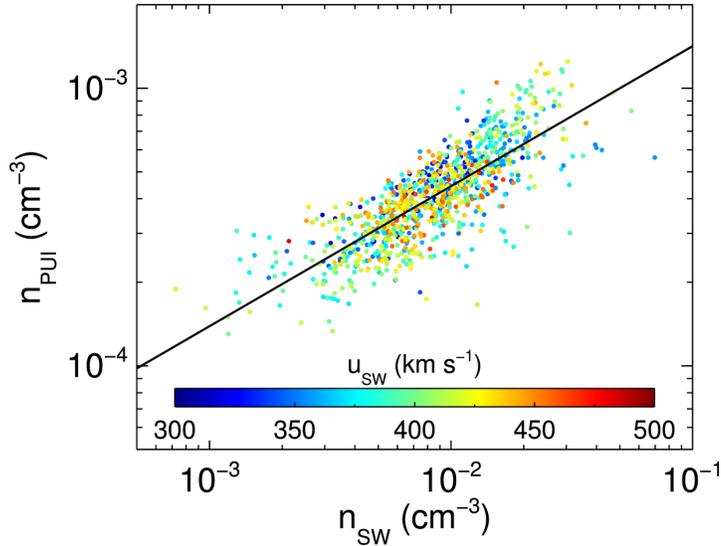

**Figure 9.** *Relationship between $H^+$ pickup ion density and solar wind core density. The strong correlation is seen at all solar wind speeds (color coded).*

The best fit power law for the correlation shown in Figure 9 is $n_{PUI} = 0.0045\, n_{SW}^{0.50}$. That is, the pickup H+ density, $n_{PUI}$, is almost exactly proportional to the square root of the solar wind density. The slower drop-off in pickup ion density makes good sense qualitatively as the pickup ions are picked up at all distances and thus represent an integrated effect compared to the local solar wind density. Finally, it is interesting there does not appear to be a strong ordering by the solar wind speed (color coded), probably because speed differences have largely worn down by these distances and primarily what is left are regions of enhanced temperature and pressure (Elliott et al. 2016).

**4. DOES THE VASYLIUNAS & SISCOE MODEL REALLY WORK?**

It seems at first remarkable that something as simple as the V&S model could provide such a good fit of the pickup hydrogen observations for the vast majority of the $H^+$ pickup ion spectra all the way out to distances of ~40 AU. This could indicate that very rapid angular scattering from a pickup ring onto an isotropic shell, followed by convection and adiabatic cooling really does a great job of characterizing the development and evolution pickup ion distributions to these distances. However, the story may also be more complicated (and interesting).

Throughout this section, we use $\beta_{HF}$ and $\lambda_{HF}$, as fitting parameters (note "F" subscript) for the H ionization rate at 1 AU and H ionization cavity scale size, respectively. However, the true H ionization rate $\beta_H$ is independently obtainable by calculating the rates of charge exchange with solar wind ions and photoionization by solar EUV emissions. Because both methods of ionization vary significantly over quite short timescales (less than a day), the $\beta_H$ at various locations in the solar wind can vary significantly and typically resides in a range normalized to 1 AU of ~$5 \times 10^{-7}$ to $1 \times 10^{-6}$ s$^{-1}$ (e.g., see McComas et al. 2017b, Figure 31).

In contrast to an ionization rate that can vary quickly in time, the H ionization cavity, $\lambda_H$, should be a much more stable value and only vary over longer timescales, such as prolonged changes in the solar wind outflow and solar cycle. McComas et al. (1999) integrated the charge exchange rate, which is the dominant ionization rate for interstellar H, with solar wind proton observations directly from the Ulysses spacecraft. These authors showed a latitude-dependent shape of the H ionization cavity for the first time and found a 1/e scale size for $\lambda_H$ of ~ 4.5 AU toward the upwind direction. Based on simulations of the distribution of interstellar neutral H density in the heliosphere with a time-dependent radiation pressure and ionization rate, Bzowski et al. (2001) showed that $\lambda_H$ varies during the solar cycle ±10%. Because New Horizons' outward trajectory is currently ~30° from this direction and because the ionization varies over time, values in the range of 3-6 AU are reasonable for $\lambda_H$.

Figure 10 shows the derived $\beta_{HF}$ and $\lambda_{HF}$ for all data points used in the radial trend analyses in this study. We note that the general relationship between these parameters is as expected since a higher real ionization rate, $\beta_H$, for a long enough time would not allow as many interstellar H neutrals to penetrate as close in to the Sun and thus produces a larger real ionization cavity scale size, $\lambda_H$. In fact, in a simple steady state V&S model, there is a relationship between the two: $\lambda_H = \beta_H v_{LISM}^{-1} r_E^2$, where $v_{LISM}$ is the interstellar bulk speed (~25 km s$^{-1}$), and $r_E$ is the distance from the Sun where $\beta_H$ is normalized. However, clearly the fitting parameters $\beta_{HF}$ and $\lambda_{HF}$ span much broader ranges of values than normally expected for the physical parameters $\beta_H$ and $\lambda_H$; in fact, only a tiny fraction (~4%) fall within the range of 1 AU scaled nominal values from this equation (red box).

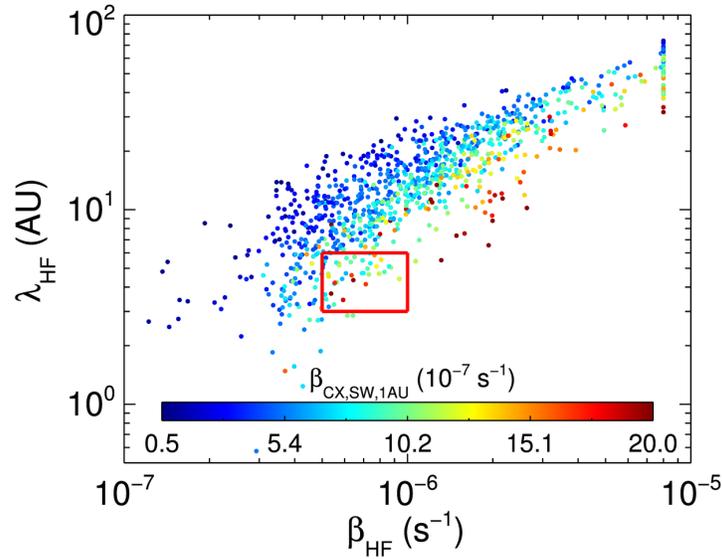

**Figure 10.** *Fit parameters for the one AU-scaled H ionization rate, $\beta_{HF}$, and the H ionization cavity, $\lambda_{HF}$, derived from minimizing the difference between the SWAP observations and model. The red box indicates the nominal range of 1 AU scaled $\beta_H$ (5-10x$10^{-7}$ s$^{-1}$) and $\lambda_H$ (3-6 AU).*

In Figure 10, we color code the calculated charge exchange rate, scaled to 1 AU, using the H-H$^+$ charge exchange cross section and solar wind speed and density values measured by SWAP. This rate clearly orders the combinations of fit parameters, with lower ionization rates being associated with smaller $\beta_{HF}$, which is not surprising. Further, the shape of the fit function tends toward a lower $\beta_{HF}$ when the data is higher density as the fit is driven by the highest 3 data points. The very large, and unphysical range of both $\beta_{HF}$ and $\lambda_{HF}$ leads us to conclude that while the general shape of the V&S function is good for fitting the pickup ion distributions observed by SWAP, it is not appropriate to ascribe the physical meaning originally used by V&S to these fitting result parameters.

Figure 11 provides additional information about the fits in our analysis. While the model is clearly not perfect (reduced chi-squared ~1), overall, the fitting is quite good, with ~63% of all reduced chi-squared values less than 5. However, the poorer fits are not uniformly distributed, and preferentially occur for the more unphysical portion of the $\beta_{HF}$-$\lambda_{HF}$ values, especially where derived $\beta_{HF} > 10^{-6}$ s$^{-1}$. In contrast, the better fits generally occur at more physical ionizations rates where $\beta_{HF} < 10^{-6}$ s$^{-1}$; however, the fit ionization cavity scale size $\lambda_{HF}$ is still generally several times larger than the physically meaningful range for $\lambda_{HF}$ of ~3-6 AU.

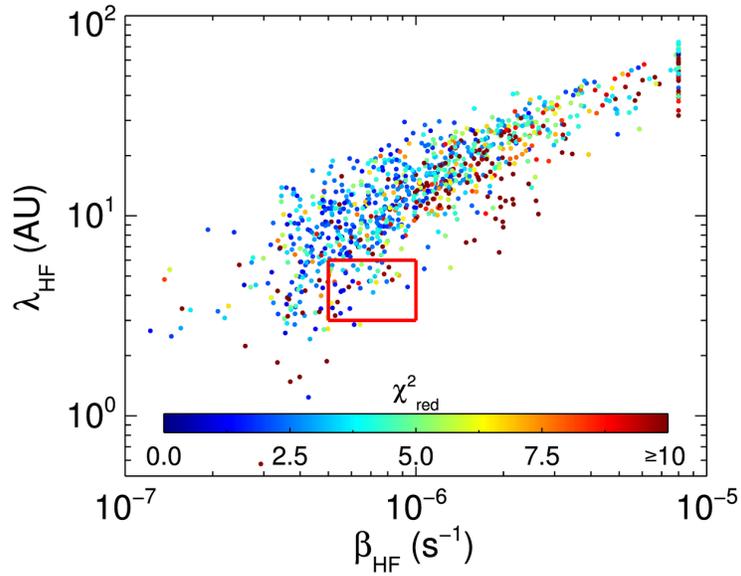

*Figure 11. Similar to Figure 10, but with color coding of the reduced chi-squared of the fit.*

To try and determine the reason that some spectra are best fit with very large $\beta_{HF}$ - $\lambda_{HF}$ values, we examined the solar wind context for such intervals. As seen in Figure 12, large $\lambda_{HF}$ (and correlated $\beta_{HF}$) times are often associated with interplanetary shocks. (Note: because *New Horizons* does not carry a magnetometer, it is not possible to absolutely differentiate discontinuous shocks from steep wave structures. Nonetheless, we use the term "shock" here less formally to describe structures that exhibit rapid, jump-like increases in speed.) It is interesting that while the sharper, shock-like structures at ~34.43, 34.63, 34.96, and 35.25 AU in the speed are associated with narrower peaks in $\lambda_{HF}$, the broader speed increase from ~35.7-36 AU is associated with a broad peak in $\lambda_{HF}$.

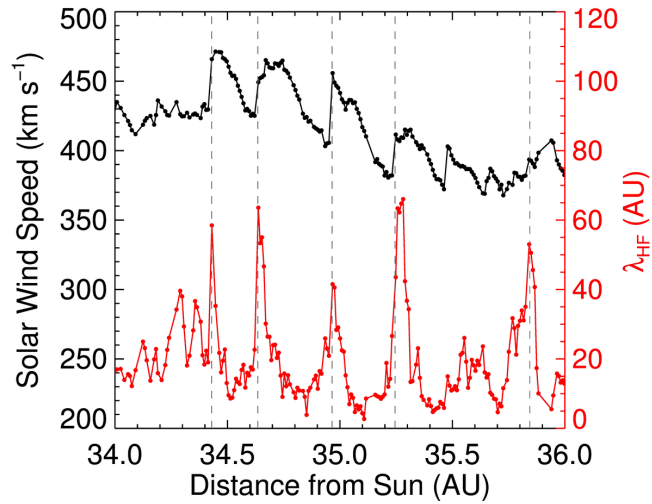

***Figure 12***. *Correlation between increases in the solar wind speed and very large values of $\lambda_{HF}$. This interval from 34-36 AU is generally characteristic of this correlation, although some speed enhancements do not show similar peaks in $\lambda_{HF}$.*

To assess the effect of the large $\beta_{HF}$-$\lambda_{HF}$ values, which seem to be poorer fits and miss some of observed distributions, for the statistical results in Section 3, we recalculated values in Table 1 after removing the highest value data points. For example, when we remove all points with $\lambda_{HF}$>40 (79 of the 987 data points or 8%), the pickup ion-to-solar wind pressure ratio at 90 AU drops from ~350 to ~300, and when we remove $\lambda_{HF}$> 30 (an additional 54 points or 5%), the pickup ion-to-SW pressure ratio at 90 AU drops to ~270. Thus, we conclude that while some of the fits become worse at times of interplanetary shocks and speed increases, and values of the extrapolations vary somewhat, the overall trends and results of this analysis remain intact.

Why would the V&S function provide good fits to the observed SWAP spectra even if the values of $\beta_{HF}$ and $\lambda_{HF}$ were very different from their physically-derived values? We believe that this is because pickup ion distributions can still be reasonably characterized by three fundamental attributes: 1) an overall scaling of the function (provided by $\beta_{HF}$), 2) an exponential energy relationship (scaled by $\lambda_{HF}$), and 3) a sharp cutoff at twice the solar wind speed (obtained by the Heaviside step function). We find that to the extent that pickup ion distributions from 20-40 AU can still be characterized by these three empirical properties, the V&S functional form is still useful, even if $\beta_{HF}$ and $\lambda_{HF}$ are just used as fitting parameters.

Given the unphysical nature of $\beta_{HF}$ and $\lambda_{HF}$, it is worth asking if the quantitative values for pickup ion density, temperature, and pressure derived above and examined for radial trends in Section 3 are still valid. We argue that they are. The fitting curves match the observed data points well and the "moment" parameters are just measures of the total amount under the curve (density), width or second moment of the distribution (a quantity like temperature), and their combination, which is the suprathermal particle pressure.

## 5. ADDITIONAL STRUCTURE IN OBSERVED PICKUP ION SPECTRA

The majority of the SWAP pickup ion spectra examined in this study can be well described by a V&S model function. In this section, we document several other types of unusual spectra that do not fit this functional form as well.

### 5.1 Tails at energies above the H pickup cut off

The SWAP spectra show a variety of enhancements or "suprathermal tails" in the H$^+$ pickup ions above the nominal cutoff at twice the solar wind peak speed (four times the E/q). Such suprathermal ion tails have been observed previously with SWICS and other instruments (Gloeckler et al. 1994; Schwadron et al. 1996; Gloeckler 1999; Hill &

Hamilton 2010; Popecki et al. 2013; Wimmer-Schweingruber et al. 2013) and, in one study, were found to correlate with compressive magnetosonic waves likely arising from statistical acceleration through transit time damping in the slow solar wind (Schwadron et al., 1996). However, there is no clear consensus as to the origin of these tails (Desai et al. 2010; Jokipii & Lee, 2010; Mason & Gloeckler, 2012; Antecki et al. 2013). Figure 13 shows an example of a particularly large enhancement of the distribution between the $H^+$ PUI cutoff and the $He^+$ PUI ledge. Clearly there are many ions at higher energies in such an energetic tail, which means that the additional pressure in this part of the distribution can constitute a large fraction of the total pressure. Significant tails are relatively unusual in the SWAP observations. However, because these ions and their pressure is explicitly not captured in the V&S fitting model used here, when they occur, the results from our fitting underestimate the real density, temperature, and pressure of such pickup ion distributions.

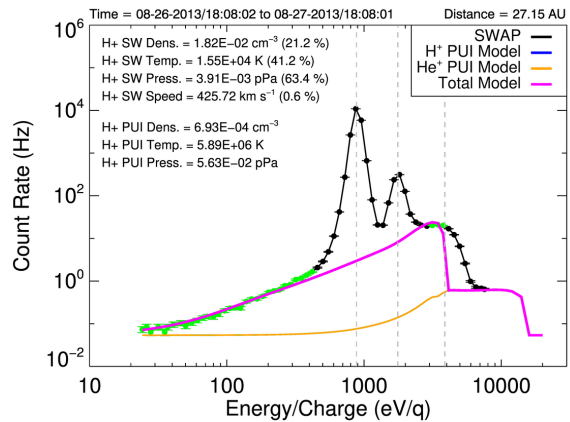

*Figure 13. Spectrum showing a significant enhanced suprathermal tail of $H^+$ pickup ions above the pickup cutoff energy. Cases like these are not included in the radial trend results in Sections 3 and 4.*

**5.2 Enhanced fluxes at energies just below H Pickup cut off**

A fascinating feature in a small fraction of the SWAP spectra is a significant enhancement at energies just below $H^+$ cutoff. Figure 14 shows an example of this in a spectrum taken on December 10-11, 2016, when New Horizons was at ~37. These spectra were among a very small fraction (~2%) manually rejected from the statistical database of derived parameters because the V&S model clearly cannot match this physical distribution.

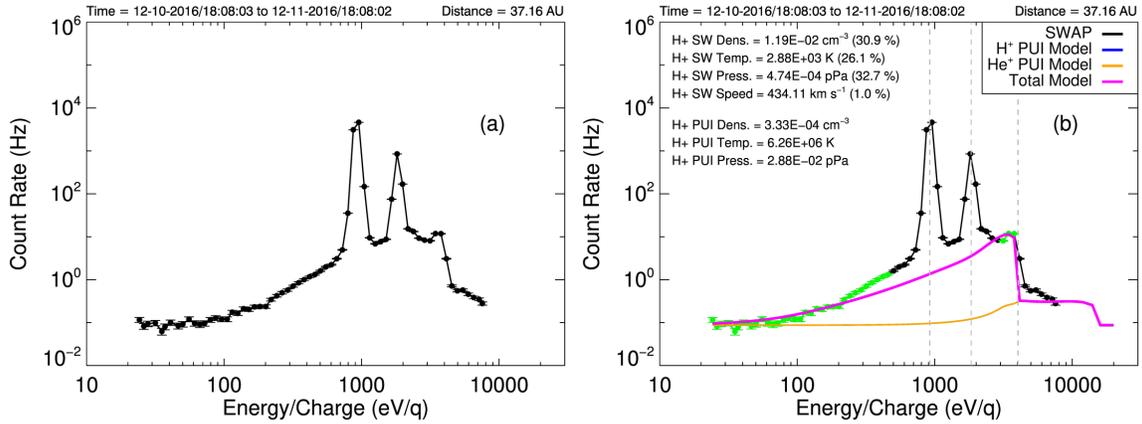

*Figure 14. (a) Spectrum showing a significant enhancement or "bump" in $H^+$ pickup ions in the two points at energies just below the cutoff energy. (b) The best fit to a V&S $H^+$ model spectrum, which has a large reduced chi-square (45.5) and clearly does not fit the observations; fits for these shaped spectra were manually rejected.*

While more work will need to be done to model this type of distribution, we suggest here that such bumps may be indicative of significantly enhanced local pickup. Significantly slower local cooling might also be possible. We note that these enhancements are sometimes associated with interplanetary shocks and the large $\beta_{HF}$-$\lambda_{HF}$ spectra (see Figure 12). Since pitch angle scattering normally occurs much more rapidly than the cooling of pickup ions, a large enhancement in charge exchange or any other sort of ionization would manifest itself as an extra enhancement of ions essentially on the pickup sphere with energies at and just below the $H^+$ cutoff energy.

### 5.3 Structure in $He^+$ pickup ion spectra above the $H^+$ pickup cut off

In this study, we have primarily focused on new observations of the pickup $H^+$ population. However, SWAP also measures $He^+$ pickup ions from above the $H^+$ pickup cutoff (and when present the $H^+$ suprathermal tail) up to energies of 7.8 keV/q (FWHM). This means that while the shape of the $H^+$ pickup ion cut off is always measured at these heliocentric distances, the cut off energy for the pickup $He^+$ is not observed. Instead, only a fraction of the distribution above the $H^+$ cut off is seen in each spectrum. That fraction varies with the solar wind speed, with more of the $He^+$ distribution exposed the slower the solar wind.

Usually, the measured portion of the $He^+$ spectra are quite flat. This generally makes sense for a V&S-shaped spectra because the $\lambda_{He}$ is so much smaller (in this study we use a fixed 0.5 AU) than the heliocentric distances of our observations (~20-40 AU). Under these conditions, the V&S spectrum becomes filled in, producing a flat "ledge," with roughly steady count rates over a broad range of higher energies. These model spectra are shown by the orange curve (covered by purple, total, curve above the $H^+$ cutoff) in Figure 4 and similar plots in this study. Even though we don't attempt to fit to the $He^+$ portion of the spectra, in general, we find amplitudes of the He$^+$ ledge that are close to those from the simple model used. In some cases, however, of the spectra do have higher and lower values, suggesting that other parameter choices, such as the rate of ionization or cavity size might better fit those samples.

In the SWAP observations, we also see a broad variety of significant deviations from this sort of flat He$^+$ distribution. In particular, we find enhancements in He$^+$ above H$^+$ cutoff and varying shapes of the He$^+$ distribution with flat portions and upward and downward steps and slopes; several examples are shown in Figure 15. These sorts of structures suggest the possibility of variations in the pickup ion history and processing.

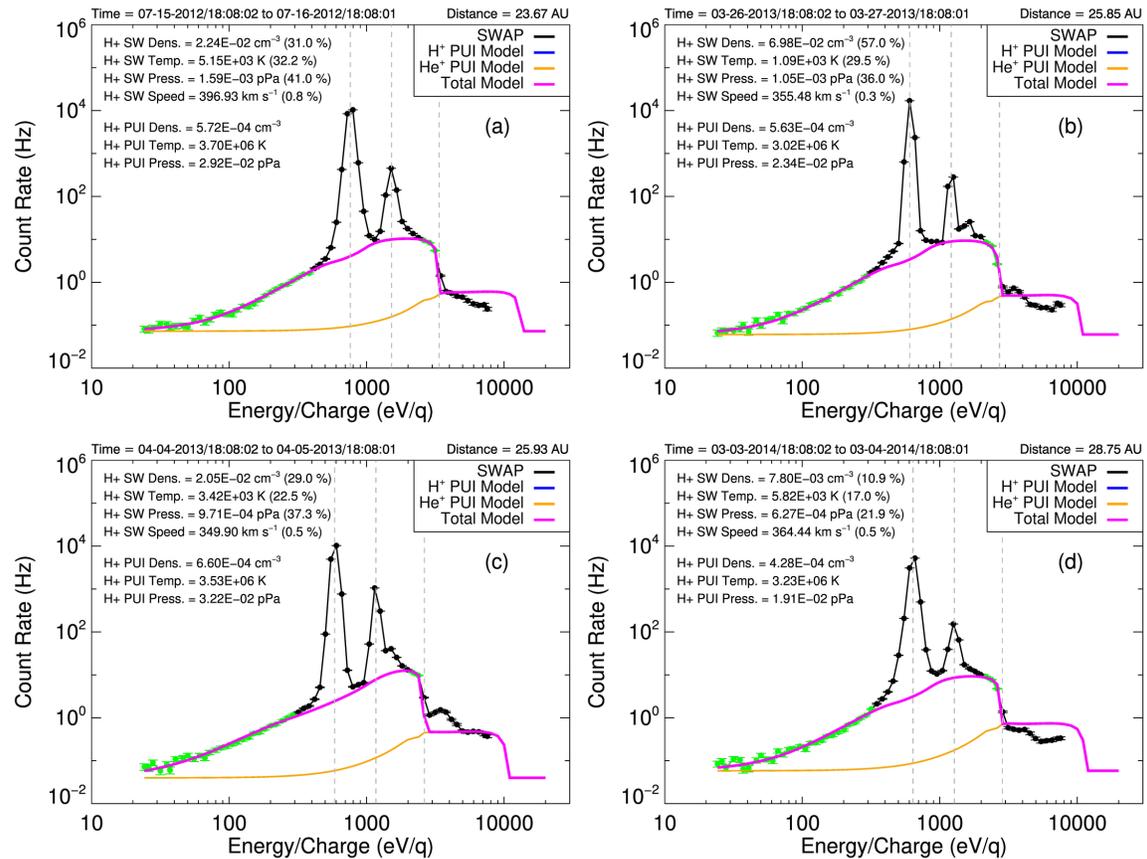

*Figure 15. Examples of spectra were the He$^+$ pickup ions don't exhibit relatively flat "ledges" above the H$^+$ cutoff, as they usually do during quiet-time solar wind.*

When the solar wind is slower, we see a larger fraction of the He$^+$ pickup ion ledge and at times these even extend up close to where the He$^+$ cutoff is. Frequently when these conditions exist, we observe a bump in the spectrum at the few energy steps just below this cutoff. This shape is highly reminiscent of the bumps observed just below the H$^+$ pickup ion cutoff (section 5.2), although He$^+$ pickup ion bumps seem to occur a larger fraction of time when the solar wind is slow enough to observe them. Figure 16 shows an example from 11/16/14, when New Horizons was at 30.9 AU. For this case, the solar wind speed was only ~ 320 km s$^{-1}$, so the highest E/q SWAP observations at ~7.8 keV/q nearly reaches the He+ cutoff.

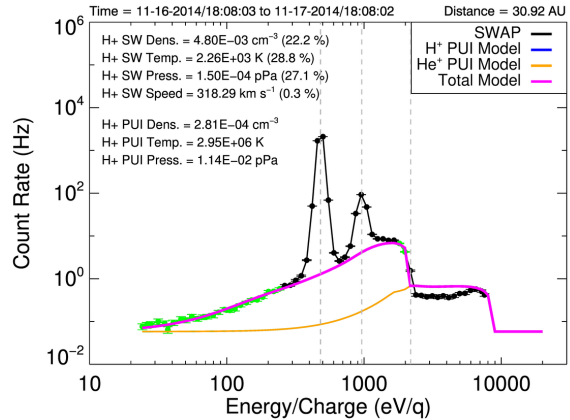

*Figure 16. SWAP observations in the slow solar wind, when most of the He$^+$ pickup population can be seen, often show enhancements or bumps at the energies just below the He$^+$ cutoff, as seen in this example.*

The details of the pickup He+ distributions are extremely interesting and SWAP provides a tremendous amount of new information about interstellar He pickup across the range from ~20-40 AU and a broad range of solar wind conditions. For example, the bumps just below the cut-offs might tell us if the pitch-angle scattering length increases with solar distance. The fact they occur more often for He than for H might be evidence for this as it is harder to scatter the heavier ions than the lighter ones. If so, the "fresher" PUIs (i.e., injected at farther distances from the Sun) would be less readily pitch angle-scattered than those injected closer to the Sun. Detailed analysis of this and other features are left to future studies of the details of these distributions.

## 6. DISCUSSION AND CONCLUSIONS

In this study, we have observed interstellar H$^+$ and He$^+$ pickup ions from 22-38 AU for the first time and:

- Measured the pickup H$^+$ moments and their radial variations
- Extrapolated observed pickup ion moments to the termination shock and compared these results to a simple mass, momentum, and energy conservation model
- Shown that H$^+$ pickup ions likely receive additional heating and that temperatures and pressures at the termination shock are larger than previously expected
- Demonstrated that the V&S model can be used to fit most of these pickup spectra, but not necessarily with physically meaningful parameters
- Shown suprathermal H$^+$ tails and multiple unusual structures in He$^+$ pickup ions
- Discovered enhancements in both H$^+$ and He$^+$ spectra just below their cutoffs

- Provided the community with unprecedented and well documented pickup ion observations for comparison to other models and theories

By fitting SWAP's observed solar wind and pickup ion data from 22-38 AU we provide characteristic values at ~30 AU (Table 1) and find the ratios of density (~0.04), temperature (~500), and particle pressure (~200) for the pickup $H^+$ compared to core solar wind protons at this distance. We also extract the radial trends of the various core solar wind and pickup ion plasma parameters and extrapolate them to the termination shock at ~90 AU, assuming they follow power laws with heliocentric distance. Because the New Horizons trajectory is headed nearly in the upwind direction (see Figure 1), the extrapolated values are indicative of the termination shock near the nose of the heliosphere. Perhaps most surprising is that the pickup ion temperature and pressure generally increase with distance from 22-38 AU, suggesting additional heating of the pickup ions.

The simple model provided above, which assumes a steady radial expansion of the solar wind and conserves mass, momentum, and energy, gave consistent values for the various parameters, except for the pickup ion temperature and internal pressure. For those, the model showed decreasing temperature and pressure with distance, in contrast to the slight increases observed here. This increase with distance in the observations produced a ratio of pickup ion pressure to solar wind dynamic pressure of 16% - over two and a half times that of the model calculation. Such a large pickup ion pressure is sufficient to significantly moderate the termination shock, as suggested by the Voyager observations.

As mentioned above, the SWAP observations may be telling us that there is some form of additional heating of the pickup ions at these distances. One possibility is heating from compression of plasma driven by the faster solar wind parcels overtaking slower parcels. In the inner heliosphere, this process produces CIRs, which then pile up into even larger MIRs in the outer heliosphere. While the speed differences eventually wear down with distance, the plasma is left with regions of higher density and temperatures that persist as the plasma continues to flow out to the termination shock.

One other interesting aspect of the model comparison comes from the fact that the model has only one significant free parameter: the interstellar neutral H density at the termination shock. The other solar wind parameters of speed and density at 30 AU are directly measured by SWAP, so the assumed neutral H density controls the modeled pickup $H^+$ density and pressure. The fact that the model appears to match observations best for a relatively low value of the interstellar neutral H density of 0.075 cm$^{-3}$ is interesting, but of course the V&S model is extremely simplified. In the future, more complete modeling of the inflowing neutral He measured by IBEX, combined with these new observations of pickup $He^+$ from SWAP, could provide an improved measure of the critical value of the upstream interstellar density. The next steps are for the community to compare the broad set of new pickup ion observations provided in this study with more complete and complex models, and in fact to develop even better new models.

This work and previous SWAP PUI observations (McComas et al. 2010; Randol et al. 2012, 2013) that have assumed an isotropic PUI model do not necessarily conflict with Ulysses observations, some of which shown large anisotropies (e.g., Gloeckler et al. 1995). Fisk et al. (1997) showed evidence from Ulysses data indicating that pickup ion distributions appears to be roughly uniform in pitch-angle hemispheres where the sunward and anti-sunward hemispheres are separated by 90° pitch-angle. Subsequent work (Isenberg 1997; Schwadron 1998; Lu & Zank 2001) developed analytical and numerical approaches to treat such hemispheric distributions. However, significant pitch-angle anisotropy should only occur when the background magnetic field is quasi-radial, and while the magnetic field is not observed on New Horizons, other observations support this caveat (Mobius et al. 1998; Oka et al. 2002; Saul et al. 2007; Drews et al. 2015, 2016). An important controlling parameter on the anisotropy appears to be the overall level of magnetic field turbulence (Oka et al. 2002; Saul et al. 2007). Since New Horizons does not measure the local magnetic field, estimation of any these parameters is not possible.

Another approach would be to utilize new computational tools such as the Energetic Particle Radiation Environment Module (EPREM, Schwadron et al., 2010, 2015), which is a 3-D kinetic model of the acceleration and transport of energetic particles along and across magnetic fields. EPREM numerically solves the focused transport equation (Skilling 1971, Ruffolo 1995; Kóta et al. 2005) and convection-diffusion equation (Jokipii et al. 1977; Lee & Fisk 1981). It has been coupled successfully to 3-D MHD codes: BATSRUS (see Toth et al., 2005, 2012), as done in Kozarev et al. (2013), and MAS (see Linker et al. 1999; Mikic et al. 1999) as done in Schwadron et al. (2015). Pickup ions are introduced into EPREM through a source term, and the model follows the propagation and acceleration of these populations within compressions, at shocks, and due to second-order mechanisms (e.g., Chen et al. 2015). EPREM can be used to model particle acceleration to suprathermal energies through the continual sampling of the velocity gradient (as in a CIR or GMIR) or from stochastic compressions. The transport of helium pickup ions in a CIR has been simulated with EPREM and has been shown to accelerate the pickup ions to suprathermal energies (Chen et al. 2015; see Figure 4).

Other areas for more detailed analysis of the SWAP observations include 1) examining the individual daily spectra around various interplanetary shocks in order to better identify the specific effects on the pickup distributions from such changes in the solar wind, 2) cataloging and determining which solar wind conditions produce the observed suprathermal tails and what physical processes drive them, and 3) understanding the detailed processes and conditions that lead to the bump enhancements in the energies just below the $H^+$ cutoff (occasionally) and $He^+$ cutoff (often, at least at slow solar wind times when we can observe it). These and many other detailed analyses, theories, and model comparisons can now be carried out to significantly advance the state of understanding of pickup ions in the solar wind.

Observations provided in this study have profound implications for the interaction of the heliosphere with the very local interstellar medium and the ongoing IBEX and Voyager missions. They are also critical for the upcoming Interstellar Mapping and Acceleration Probe (IMAP) mission, which will examine this interaction much more broadly. Extending the measurements by SWAP on New Horizons over the next decade would continue to provide critical new information about the pickup ions through a region of the heliosphere where they play an increasing critical role and where no such measurements have ever been taken.

This study has analyzed and documented the SWAP pickup ion observations from 22 to beyond 38 AU for the first time and serves as the citable reference for use of the SWAP observations by broader scientific community. These data can be accessed through the (NOTE: LINK will be added once the paper is accepted).

*Acknowledgements*. We gratefully thank all the SWAP and New Horizons team members who have made this instrument and mission possible. As per the requirements of IOP policy on attributions of credit and responsibility in authorship, we are disclosing that S.A. Stern, C. Olkin, J. Spencer, and H. Weaver were added as coauthors at the direction of the New Horizons mission PI because "the PI and project scientists do the work that makes all the papers possible, from 2001 onward. Without our contributions, this paper could not exist." They are not to be held responsible for the scientific observations, analysis, or results of this study. This work was carried out as a part of the SWAP instrument effort on the New Horizons project, with support from NASA's New Frontiers Program and the Polish National Science Center grant 2015/18/M/ST9/00036.